\documentclass[aps,prl,twocolumn,showpacs]{revtex4}

\usepackage{epsf}

\begin{document}

\title{Surface electromagnetic phenomena in pristine and atomically doped carbon nanotubes}

\author{I.V.~Bondarev}\email[Corresponding author.
E-mail: ]{ibondarev@nccu.edu}\affiliation{Physics Department,
North Carolina Central University, 1801 Fayetteville Str, Durham,
NC 27707, USA}

\begin{abstract}
The article reviews recent progress in the theoretical
understanding of near-field surface electromagnetic phenomena in
pristine and atomically doped carbon nanotubes. The phenomena
involving strong coupling effects are outlined. They are the
optical absorption by single-walled carbon nanotubes doped with
single atoms or ions in the frequency range close to the atomic
transition frequency, the entanglement of the pair of atomic
qubits strongly coupled to a common high-finesse surface photonic
mode of the nanotube, and the optical response of the strongly
coupled surface exciton-plasmon excitations in pristine
semiconducting carbon nanotubes. The phenomena reviewed have a
great potential to be exploited for the future development of the
nanotube based tunable optoelectronic device applications in areas
such as nanophotonics, nanoplasmonics, cavity quantum
electrodynamics, and quantum information science.
\end{abstract}
\pacs{78.40.Ri, 73.22.-f, 73.63.Fg, 78.67.Ch}

\maketitle

\newpage

\noindent\textbf{1. Introduction}

Single-walled carbon nanotubes (CNs) are quasi-one-dimensional
(1D) cylindrical wires consisting of graphene sheets rolled-up
into cylinders with diameters $\sim\!1-10$~nm and lengths
$\sim\!1-10^4\,~\mu$m~\cite{Dresselhaus,Dai,Zheng,Huang}.
Nanotubes are shown to be useful as miniaturized electronic,
electromechanical, and chemical devices~\cite{Baughman}, scanning
probe devices~\cite{Popescu}, and nanomaterials for macroscopic
composites~\cite{Trancik}. Their intrinsic physical properties may
be substantially modified in a controllable way by doping with
extrinsic impurity atoms, molecules and
compounds~\cite{Latil,Son,Duclaux,Shimoda,Jeong,Jeongtsf,Khazaei}.

Recent successful experiments on the encapsulation of single atoms
into single-walled CNs~\cite{Jeong,Jeongtsf,Khazaei} and their
intercalation into single-wall CN bundles~\cite{Duclaux,Shimoda},
along with the numerous previous studies of monoatomic gas
absorption by the nanotube bundles (see, e.g., Ref.~\cite{Calbi})
and the progress in growth techniques of centimeter-long
small-diameter single-walled carbon nanotubes~\cite{Zheng,Huang},
stimulated an in-depth theoretical analysis of dynamic quantum
coherent processes in CNs doped with single atoms or ions. As a
result, atomically doped nanotubes have been recently demonstrated
to be very efficient as applied to nanophotonics, cavity quantum
electrodynamics (QED) and quantum
communication~\cite{Bondarev06,Bondarev07,Bondarev07jem,Bondarev07os}.

For pristine (undoped) single-walled CNs, the numerical
calculations predicting large exciton binding energies
($\sim\!0.3\!-\!0.6$~eV) in semiconducting
CNs~\cite{Pedersen03,Pedersen04,Capaz}~and even in some
small-diameter ($\sim\!0.5$~nm) metallic CNs~\cite{Spataru04},
followed by the results of various exciton photoluminescence
measurements~\cite{Imamoglu,Wang04,Wang05,Hagen05,Plentz05,Avouris08},
have become available.~These works, together with other reports
investigating the role of effects such as intrinsic
defects~\cite{Hagen05,Prezhdo08}, exciton-phonon
interactions~\cite{Plentz05,Prezhdo08,Perebeinos05,Lazzeri05,Piscanec},
exciton-surface-plasmon coupling~\cite{Bondarev08,Bondarev09},
effects of external magnetic~\cite{Zaric,Srivastava} and electric
fields~\cite{Perebeinos07}, reveal the variety and complexity of
the intrinsic optical properties of semiconductor carbon
nanotubes~\cite{Dresselhaus07}. At the same time, they pave the
way for the novel applications of carbon nanotubes in modern
optoelectronics in areas such as various aspects of photophysics,
nanoplasmonics, and quantum information science, which are
essentially beyond relatively well explored nanoelectronics and
nanoelectromechanics with carbon nanotubes. For instance, recent
experimental observations of quantum correlations in the
photoluminescence spectra of individual CNs suggest their
potential applicability in quantum cryptography~\cite{Imamoglu}.

In spite of impressive experimental demonstrations of basic
quantum information effects in a number of different nanoscale
solid state systems, such as quantum dots in semiconductor
microcavities, nuclear spin systems, Josephson junctions, etc.,
their concrete implementation is still at the proof-of-principle
stage (see, e.g., Ref.~\cite{Brandes} for a review). The
development of materials that may host quantum coherent states
with long coherence lifetimes is a critical research problem for
the nearest future. There is a need for the fabrication of quantum
bits (qubits) with coherence lifetimes at least three-four orders
of magnitude longer than it takes to perform a bit
flip~\cite{Brandes,Lukin,Benjamin}. To achieve this goal, a
critical prerequisite of the strong coupling of an atomic,
excitonic or intersubband electronic transition to a high-finesse
vacuum-type electromagnetic mode of its environment (e.g., a
semiconductor optical microcavity mode) needs to be fulfilled.
Under strong coupling, new elementary excitations, the eigen
states of the full photon-matter Hamiltonian, form. They are
so-called 'cavity polaritons' representing a characteristic
anticrossing behavior with a mode separation often referred to as
'vacuum-field Rabi splitting'. A pair of the cavity polaritons
(term used in a broad sense here) coupled to the same high-finesse
vacuum-type electromagnetic mode of the surrounding material would
involve polariton entangling
operations~\cite{Bondarev07,Lukin,Benjamin,Hughes,Cirac}, followed
by the nearest neighbor interaction over short distances and
quantum information transfer over longer distances.

For atomically doped carbon nanotubes, there may be typically two
qualitatively different regimes of the interaction of an atomic
state with the medium-assisted surface electromagnetic modes of a
CN~\cite{Bondarev02,Bondarev04,Bondarev04pla,Bondarev05,Bondarev04ssc,Bondarev06trends}.
They are the weak coupling regime and the strong coupling regime.
The former is characterized by the monotonic exponential decay
dynamics of the upper atomic state with the decay rate altered
compared to the free-space value due to the presence of the
additional decay modes associated with the presence of the
nanotube. The latter is, in contrast, strictly non-exponential
and~is characterized by the reversible (Rabi) oscillations where
the energy of the initially excited atom is periodically exchanged
between the atom and the surface electromagnetic mode of the
nanotube. The strong coupling regime was shown recently to occur
in small-diameter ($\sim\!1$~nm) atomically doped CNs with the
atomic transition tuned to the resonance (representing the field
mode) of the local density of the nanotube's surface photonic
states
(DOS)~\cite{Bondarev04,Bondarev04pla,Bondarev05,Bondarev04ssc,Bondarev06trends}.
This result can be qualitatively understood in terms of the cavity
QED. The coupling constant of an atom (modelled by a~two-level
system with the transition dipole moment $d_{A}$ and frequency
$\omega_{A}$) to a local vacuum-type field is given by $\hbar
g\!=\!(2\pi d_{A}^{2}\hbar\omega_{A}/\tilde{V})^{1/2}$ with
$\tilde{V}$ being the effective volume of the field mode the atom
interacts with (see, e.g., Ref.~\cite{Andreani}). For the atom
(ion) encapsulated into the CN of radius $R_{cn}$, the local field
mode volume is estimated to be $\tilde{V}\!\sim\!\pi
R_{cn}^{2}(\lambda_{A}/2)$ that is $\sim\!10^2$~nm$^3$ for CNs
with diameters $\sim\!1$~nm in the optical range of
$\lambda_{A}\!\sim\!600$~nm. Approximating
$d_{A}\!\sim\!er\!\sim\!e(e^{2}/\hbar\omega_{A})$~\cite{Davydov},
one obtains $\hbar g\!\sim\!0.3$~eV. On the other hand, the
"cavity" linewidth is given for $\omega_{A}$ in resonance with the
cavity mode by $\hbar\gamma_{c}\!=\!6\pi\hbar
c^{3}/\omega_{A}^{2}\xi(\omega_{A})\tilde{V}$, where $\xi$ is the
atomic spontaneous emission enhancement/dehancement (Purcell)
factor~\cite{Andreani}. Taking into account large Purcell factors
$\sim\!10^{7}$ close to CNs~\cite{Bondarev02}, one arrives at
$\hbar\gamma_{c}\!\sim\!0.03$~eV for 1~nm-diameter CNs in the
optical spectral range. Thus, for the atoms (ions) encapsulated
into the small-diameter CNs the strong atom-field coupling
condition $g/\gamma_{c}\gg1$ is fulfilled, giving rise to the
rearrangement ("dressing") of the atomic levels followed by the
formation of quasi-one-dimensional (1D) atomic cavity
polaritons~\cite{Bondarev04,Bondarev04pla,Bondarev05,Bondarev04ssc,Bondarev06trends}.
The latter ones are similar to quasi-0D excitonic polaritons in
quantum dots in semiconductor
microcavities~\cite{Reithmaier,Yoshie,Peter}, which are
theoretically proposed to be a way to produce the excitonic states
entanglement for use in solid state quantum information science
applications~\cite{Hughes}. Similar way to entangle the pair of
spatially separated quasi-1D atomic polaritons in metallic
small-diameter single-walled CNs has been theoretically
demonstrated to be feasible as well~\cite{Bondarev07}. It is
important to pursue a variety of different strategies and
approaches towards physically implementing novel non-trivial
applications in modern nanotechnology.

For pristine small-diameter ($\lesssim\!1$~nm) semiconducting
single-walled CNs, the exciton-surface-plasmon interactions have
been shown recently to result in the strong exciton-plasmon
coupling regime~\cite{Bondarev08,Bondarev09}. This is due to the
fact that the low-energy weakly-dispersive interband plasmon modes
(resulted from the transverse quantization of the electronic
motion on the nanotube surface --- observed in
Ref.~\cite{Pichler98}) occur in the same energy range of
$\sim\!~1$~eV where the lowest exciton excitation energies are
located in small-diameter CNs~\cite{Spataru05,Ma}. Previous
studies of the exciton-plasmon interactions in nanomaterials have
been focused on artificially fabricated \emph{hybrid} plasmonic
nanostructures, such as dye molecules in organic polymers
deposited on metallic films~\cite{Bellessa}, semiconductor quantum
dots coupled to metallic nanoparticles~\cite{Govorov}, or
nanowires~\cite{Fedutik}, where one material carries the exciton
and another one carries the plasmon. The effect of the strong
exciton-plasmon coupling in CNs is particularly interesting since
here the fundamental electromagnetic phenomenon of strong coupling
occurs in an \emph{individual} quasi-1D nanostructure, a~carbon
nanotube.

This article briefly reviews recent progress in the theoretical
understanding of near-field surface electromagnetic phenomena in
pristine and atomically doped CNs. It focuses on the phenomena
that involve the strong coupling effects. Specifically, light
absorption by atomically doped CNs and the entanglement of
spatially separated strongly coupled quasi-1D atomic polaritons
are discussed. Also, physical reasons for the
exciton-surface-plasmon coupling in pristine small-diameter
semiconducting CNs is given, and the optical response of the
strongly coupled surface exciton-plasmon excitations is analyzed.
These strong coupling phenomena have a great potential to be
exploited for the future development of the nanotube based tunable
optoelectronic device applications in areas such as nanophotonics,
nanoplasmonics, cavity QED, and quantum information science.

The macroscopic electromagnetic Green function formalism is used
throughout the paper, developed in
Refs.~\cite{Bondarev02,Bondarev04,Bondarev04pla,Bondarev04ssc,Bondarev05,Bondarev06trends}
to study the quantum nature of the near-field electromagnetic
interactions in quasi-1D nanostructures featuring strong
absorption. The formalism follows the original line of the
macroscopic QED approach introduced by Welsch and coworkers to
rigorously describe medium-assisted electromagnetic vacuum effects
in dispersing and absorbing
media~\cite{VogelWelsch,ScheelWelsch,BuhmannWelsch} (and refs.
therein).

The paper is organized as follows. Section~2 analyzes the optical
absorption by single-walled CNs doped with single atoms (or ions)
in the frequency range close to the atomic transition frequency. A
significant line (Rabi) splitting effect is demonstrated for
small-diameter ($\sim\!1$~nm) CNs~\cite{Bondarev06}, which
indicates the strong atom-field coupling regime known to
facilitate the entanglement of spatially separated atomic
qubits~\cite{Cirac,Dung}. Section~3 calculates the entanglement of
formation for the pair of atomic qubits (quasi-1D atomic
polaritons) strongly coupled to the same medium-assisted
high-finesse surface photon mode of the nanotube. Small-diameter
metallic nanotubes are shown to result in sizable amounts of the
two-qubit atomic entanglement for sufficiently long
times~\cite{Bondarev07}. Section~4 describes the physical nature
of the interactions between excitonic states and interband surface
plasmons in pristine small-diameter semiconducting single-walled
CNs, and calculates the optical response (exciton absorption)
under strong exciton-plasmon coupling. Section~5 discusses the
envisaged nanotube based applications of the phenomena considered,
summarizes and concludes the article.

~

\noindent\textbf{2. Optical Absorbtion by Atomically Doped CNs}

~

\begin{figure}[t]
\epsfxsize=8.65cm\centering{\epsfbox{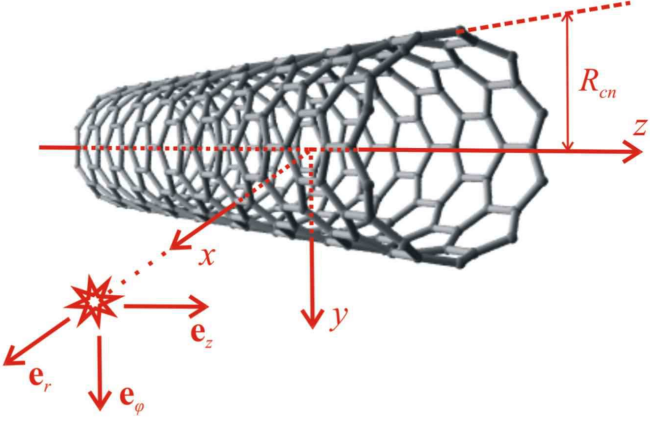}}\caption{(Color
online) The geometry of the problem.}\label{fig1}
\end{figure}

The (achiral) CN is modelled by an infinitely long, infinitely
thin, anisotropically conducting cylinder. Its (axial) surface
conductivity is taken to be that given by the $\pi$-electron band
structure in the tight-binding approximation with the azimuthal
electron momentum quantization and axial electron momentum
relaxation taken into account. A two-level atom is positioned at
the point $\mathbf{r}_{A}$ near an infinitely long achiral
single-wall~CN (Fig.~\ref{fig1}). The orthonormal cylindric basis
$\{\mathbf{e}_{r},\mathbf{e}_{\varphi},\mathbf{e}_{z}\}$ is chosen
in such a way that $\mathbf{e}_{z}$ is directed along the nanotube
axis and, without loss of generality,
$\mathbf{r}_{A}\!=\!r_{A}\mathbf{e}_{r}\!=\!\{r_{A},0,0\}$. The
atom interacts with the quantum electromagnetic field via its
transition dipole moment directed along the CN axis,
$\mathbf{d}_{A}\!=\!d_{z}\textbf{e}_{z}$. The contribution of the
transverse dipole moment orientations is suppressed because of the
strong depolarization of the transverse field in an isolated
CN~\cite{Benedict,Tasaki,Li,Marinop,Ando,Kozinsky}. Under these
assumptions, the photon Green function formalism developed earlier
in Refs.~\cite{Bondarev05,Bondarev06trends} for quantizing an
electromagnetic field close to quasi-1D absorbing and dispersing
media  generates the total second quantized Hamiltonian (Gaussian
units)
\begin{eqnarray}
\hat{H}=\int_{0}^{\infty}\!\!\!\!\!d\omega\,\hbar\omega\!\int\!d\mathbf{R}
\,\hat{f}^{\dag}(\mathbf{R},\omega)\hat{f}(\mathbf{R},\omega)+
{\hbar\tilde{\omega}_{A}\over{2}}\,\hat{\sigma}_{z}\label{Htwolev}\\[0.2cm]
+\int_{0}^{\infty}\!\!\!\!\!d\omega\!\int\!d\mathbf{R}\;[\,
\mbox{g}^{(+)}(\mathbf{r}_{A},\mathbf{R},\omega)\,\hat{\sigma}^{\dag}\nonumber\hskip1.2cm\\[0.2cm]
-\,\mbox{g}^{(-)}(\mathbf{r}_{A},\mathbf{R},\omega)\,\hat{\sigma}\,]\,
\hat{f}(\mathbf{R},\omega)+\mbox{h.c.},\nonumber\hskip1cm
\end{eqnarray}
where the three terms represent the \emph{medium-assisted}
(modified by the presence of the CN) electromagnetic field, the
two-level atom and their interaction, respectively. The operators
$\hat{f}^{\dag}(\mathbf{R},\omega)$ and
$\hat{f}(\mathbf{R},\omega)$ are the scalar bosonic field
operators defined on the CN surface assigned by the radius-vector
$\mathbf{R}=\!\{R_{cn},\phi,Z\}$ with $R_{cn}$ being the radius of
the CN. These operators create and annihilate the single-quantum
bosonic-type electromagnetic medium excitation of the frequency
$\omega$ at the point $\mathbf{R}$ of the CN surface. The Pauli
operators, $\hat{\sigma}_{z}\!=\!|u\rangle\langle
u|-|l\rangle\langle l|$,\linebreak
$\hat{\sigma}\!=\!|l\rangle\langle u|$ and
$\hat{\sigma}^{\dag}\!=\!|u\rangle\langle l|$, describe the atomic
subsystem and electric dipole transitions between the two atomic
states, upper $|u\rangle$ and lower $|l\rangle$, separated by the
transition frequency~$\omega_{A}$. This (bare) frequency is
modified by the diamagnetic ($\sim\!\mathbf{A}^{2}$) atom-field
interaction yielding the new \emph{renormalized} frequency
\[
\tilde{\omega}_{A}=\omega_{A}\left[1-\frac{2}{(\hbar\omega_{A})^{2}}\!
\int_{0}^{\infty}\!\!\!\!\!d\omega\!\int\!d\mathbf{R}\,
|\mbox{g}^{\perp}(\mathbf{r}_{A},\mathbf{R},\omega)|^{2}\right]
\]
in the second term of the Hamiltonian. The dipole atom-field
interaction matrix elements
$\mbox{g}^{(\pm)}(\mathbf{r}_{A},\mathbf{R},\omega)$ are given by
$\mbox{g}^{(\pm)}\!=\!\mbox{g}^{\perp}\pm(\omega/\omega_{A})\mbox{g}^{\parallel}$,
where
\begin{eqnarray}
\mbox{g}^{\perp(\parallel)}(\mathbf{r}_{A},\mathbf{R},\omega)=
-i\frac{4\omega_{A}}{c^{2}}\;d_{z}\hskip0.5cm\nonumber\\
\times\,\sqrt{\pi\hbar\omega\,\mbox{Re}\,\sigma_{zz}(\omega)}
^{\;\,\perp(\parallel)}G_{zz}(\mathbf{r}_{A},\mathbf{R},\omega)\nonumber
\end{eqnarray}
with $^{\perp(\parallel)}G_{zz}$ being the $zz$-component of the
transverse (longitudinal) Green tensor (with respect to the first
variable) of the electromagnetic subsystem, and
$\sigma_{zz}(\omega)$ representing the CN surface axial
conductivity. The matrix elements $\mbox{g}^{\perp(\parallel)}$
have the property of
\[
\int\!d\mathbf{R}\,|\mbox{g}^{\perp(\parallel)}(\mathbf{r}_{A},\mathbf{R},\omega)|^{2}=
\frac{\hbar^2}{2\pi}\left(\frac{\omega_{A}}{\omega}\right)^{\!2}\!\Gamma_{0}(\omega)
\xi^{\perp(\parallel)}(\mathbf{r}_{A},\omega)
\] with
\[
\xi^{\perp(\parallel)}(\mathbf{r}_{A},\omega)=\frac{\mbox{Im}^{\perp(\parallel)}
G_{zz}^{\,\perp(\parallel)}(\mathbf{r}_{A},\mathbf{r}_{A},\omega)}{\mbox{Im}\,G_{zz}^{0}(\omega)}
\]
being the transverse (longitudinal) distance-dependent (local)
photonic DOS functions, and
\[
\Gamma_{0}(\omega)=\frac{8\pi\omega^{2}d_{z}^{2}}{3\hbar\,\!c^{2}}\,\mbox{Im}\,G_{zz}^{0}(\omega)
\]
representing the atomic spontaneous decay rate in vacuum, where
$\mbox{Im}\,G_{zz}^{0}(\omega)\!=\!\omega/6\pi c$ is the vacuum
imaginary Green tensor $zz$-component.

The Hamiltonian~(\ref{Htwolev}) involves only two standard
approximations. They are the electric dipole approximation and the
two-level approximation. The rotating wave approximation commonly
used is not applied, and the diamagnetic term of the atom-field
interaction is not neglected (as opposed to, e.g.,
Refs.~\cite{Bondarev02,Bondarev04,Bondarev04pla,Dung}). Quantum
electrodynamics of the two-level atom (ion) close to the CN is
thus described in terms of only two intrinsic characteristics of
the electromagnetic subsystem --- the transverse and longitudinal
local photonic DOS functions.

When the atom is initially in the upper state and the field
subsystem is in vacuum, the time-dependent wave function of the
whole system can be written as
\begin{eqnarray}
|\psi(t)\rangle=C_{\displaystyle{u}}(t)\,e^{-i(\tilde{\omega}_{A}/2)t}
|u\rangle|\{0\}\rangle\hskip1cm\label{wfunc}\\[0.2cm]
+\int\!d\mathbf{R}\!\int_{0}^{\infty}\!\!\!\!\!\!d\omega\,
C_{l}(\mathbf{R},\omega,t)\,e^{-i(\omega-\tilde{\omega}_{A}/2)t}
|l\rangle|1(\mathbf{R},\omega)\rangle,\nonumber
\end{eqnarray}
where $|\{0\}\rangle$ is the vacuum state of the field subsystem,
$|\{1(\mathbf{R},\omega)\}\rangle$ is its excited state with the
field being in the single-quantum Fock state,
$C_{\displaystyle{u}}$ and $C_{l}$ are the population probability
amplitudes of the upper state and the lower state of the whole
system, respectively. Applying the Hamiltonian~(\ref{Htwolev}) to
the wave function~(\ref{wfunc}), one has
\begin{eqnarray}
\mbox{\it\.{C}}_{\displaystyle{u}}(t)=-\frac{i}{\hbar}\!
\int_{0}^{\infty}\!\!\!\!\!\!d\omega\!\int\!d\mathbf{R}\,
\mbox{g}^{(+)}(\mathbf{r}_{A},\mathbf{R},\omega)\hskip1cm\label{popampu}\\[0.1cm]
\times\,C_{l}(\mathbf{R},\omega,t)e^{-i(\omega-\tilde{\omega}_{A})t},\hskip2cm\nonumber\\[0.1cm]
\mbox{\it\.{C}}_{l}(\mathbf{R},\omega,t)\!=-\frac{i}{\hbar}\,
[\mbox{g}^{(+)}(\mathbf{r}_{A},\mathbf{R},\omega)]^{\!\ast}\,
C_{\displaystyle{u}}(t)e^{i(\omega-\tilde{\omega}_{A})t}.
\label{popampl}
\end{eqnarray}

In terms of the probability amplitudes above, the
emission/absorbtion spectral line shape $I(\omega)$ is given by
$\int\!d\mathbf{R}\,|C_{l}(\mathbf{R},\omega,t\rightarrow\infty)|^{2}$~\cite{Heitler}.
(Obviously, the absorbtion line shape coincides with the emission
line shape if the monochromatic incident light beam is used in the
absorbtion experiment.) This, after the substitution of $C_{l}$
obtained by the integration [under initial conditions
$C_{l}(\mathbf{R},\omega,0)\!=\!0$,
$C_{\displaystyle{u}}(0)\!=\!1$] of Eq.~(\ref{popampl}), yields in
dimensionless variables
\begin{equation}
\tilde{I}(\mathbf{r}_{A},x)=\tilde{I}_{0}(\mathbf{r}_{A},\tilde{x}_{A})
\left|\int_{0}^{\infty}\!\!\!\!\!\!d\tau\,C_{\displaystyle{u}}(\tau)\,
e^{i(x-\tilde{x}_{A})\tau}\right|^{2}\!\!, \label{Ix}
\end{equation}
where
\[
\tilde{I}=\frac{2\gamma_{0}I}{\hbar}\,,\;\;\;
\tilde{\Gamma}_{0}=\frac{\hbar\Gamma_{0}}{2\gamma_{0}}\,,\;\;\;
x=\frac{\hbar\omega}{2\gamma_{0}}\,,\;\;\;\tau=\frac{2\gamma_{0}t}{\hbar}
\]
with $\gamma_{0}\!=\!2.7$~eV being the carbon nearest neighbor
hopping integral ($\hbar/2\gamma_{0}\!=\!1.22\times\!10^{-16}$~s)
appearing in the CN surface axial conductivity $\sigma_{zz}$, and
\[
\tilde{I}_{0}(\mathbf{r}_{A},x)=\frac{\tilde{\Gamma}_{0}(x)}{2\pi}
\left[\left(\frac{x_A}{x}\right)^{\!2}\xi^{\perp}(\mathbf{r}_{A},x)+\xi^{\parallel}(\mathbf{r}_{A},x)\right].
\]

The upper state population probability amplitude in Eq.~(\ref{Ix})
is given by the Volterra integral equation [obtained by
substituting the result of the formal integration of
Eq.~(\ref{popampl}) into Eq.~(\ref{popampu})] with the kernel
determined by the local photonic DOS functions
$\xi^{\perp(\parallel)}(\mathbf{r}_{A},x)$~\cite{Bondarev05,Bondarev04ssc,Bondarev06trends}.
Thus, the numerical solution is only possible for the line shape
$\tilde{I}$, strictly speaking. This, however, offers very little
physical insight into the problem of the optical absorbtion by
atomically doped CNs under different atom-field coupling regimes.
On the other hand, for those atomic transition frequencies
$\tilde{x}_{A}$ which are located in the vicinity of the resonance
frequencies $x_{r}$ of the DOS functions $\xi^{\perp(\parallel)}$
a~simple analytical approach may be applied. Specifically,
$\xi^{\perp(\parallel)}(\mathbf{r}_{A},x\!\sim\!x_{r})$ can be
approximated by the Lorentzians of the same
half-width-at-half-maxima $\delta x_{r}$, thus making it possible
to solve the integral equation for $C_{\displaystyle{u}}$
analytically to
obtain~\cite{Bondarev04,Bondarev04pla,Bondarev06trends}
\begin{eqnarray}
C_{\displaystyle{u}}(\tau)\approx\frac{1}{2}\left(\!1+\frac{\delta
x_{r}}{\sqrt{\delta\,\!x_{r}^{2}-X^{2}}}\right)e^{-\frac{1}{2}\left(\delta
x_{r}-\sqrt{\delta\,\!x_{r}^{2}-X^{2}}\right)\tau}\label{Cuapp}\\
+\frac{1}{2}\left(\!1-\frac{\delta\,\!x_{r}}{\sqrt{\delta\,\!x_{r}^{2}-X^{2}}}\right)
e^{-\frac{1}{2}\left(\delta\,\!x_{r}+\sqrt{\delta\,\!x_{r}^{2}-X^{2}}\right)\tau}\nonumber
\end{eqnarray}
with
\begin{equation}
X=\sqrt{2\delta\,\!x_{r}\tilde{\Gamma}_{0}(\tilde{x}_{A})\xi^{\perp}(\mathbf{r}_{A},\tilde{x}_{A})}\,.
\label{X}
\end{equation}
This solution is valid for
$\tilde{x}_{A}\!\approx\!x_{r}$ whatever the atom-field coupling
strength is, yielding the exponential decay of the upper state
population,
$|C_{\displaystyle{u}}(\tau)|^{2}\!\approx\!\exp[-\tilde{\Gamma}(\tilde{x}_{A})\tau]$
with the rate
$\tilde{\Gamma}\!\approx\!\tilde{\Gamma}_{0}\xi^{\perp}$, in the
weak coupling regime where $(X/\delta x_{r})^{2}\!\ll\!1$, and the
decay via damped Rabi oscillations,
$|C_{\displaystyle{u}}(\tau)|^{2}\!\approx\!\exp(-\delta
x_{r}\tau)\cos^{2}(X\tau/2)$, in the strong coupling regime where
$(X/\delta x_{r})^{2}\!\gg\!1$.

Substituting Eq.~(\ref{Cuapp}) into Eq.~(\ref{Ix}) and integrating
over~$\tau$, one arrives at the expression
\begin{equation}
\tilde{I}(\mathbf{r}_{A},x)=\frac{\tilde{I}_{0}(\mathbf{r}_{A},\tilde{x}_{A})\left[
(x-\tilde{x}_{A})^{2}+\delta x_{r}^{2}\right]}
{\left[(x-\tilde{x}_{A})^{2}-X^{2}/4\right]^{2}+\delta
x_{r}^{2}(x-\tilde{x}_{A})^{2}} \label{Iraxfinal}
\end{equation}
representing the emission/absorbtion spectral line shape for
frequencies $x\!\sim\!x_{r}\!\approx\!\tilde{x}_{A}$ regardless of
the atom-field coupling strength. The line shape (\ref{Iraxfinal})
is clearly seen to be of a symmetric two-peak structure in the
strong coupling regime where $(X/\delta x_{r})^{2}\!\gg\!1$. The
exact peak positions are
\[
x_{1,2}=\tilde{x}_{A}\pm(X/2)\sqrt{\sqrt{1+8(\delta
x_{r}/X)^{2}}-4(\delta x_{r}/X)^{2}}\;,
\]
separated from each other by $x_{1}-x_{2}\sim X$ with the Rabi
frequency $X$ given by Eq.~(\ref{X}). In the weak coupling regime
$(X/\delta x_{r})^{2}\ll1$, and $x_{1,2}$ become complex,
indicating that there are no longer peaks at these frequencies. As
this takes place, Eq.~(\ref{Iraxfinal}) is approximated with the
weak coupling condition, the fact that $x\sim\tilde{x}_{A}$ and
Eq.~(\ref{X}), to give the well-known Lorentzian
\[
\tilde{I}(x)=\frac{\tilde{I}_{0}(\tilde{x}_{A})}{(x-\tilde{x}_{A})^{2}+\tilde{\Gamma}^{2}(\tilde{x}_{A})/4}
\]
with the half-width-at-half-maximum
$\tilde{\Gamma}(\tilde{x}_{A})/2$, peaked at $x=\tilde{x}_{A}$.

\begin{figure}[t]
\epsfxsize=8.65cm\centering{\epsfbox{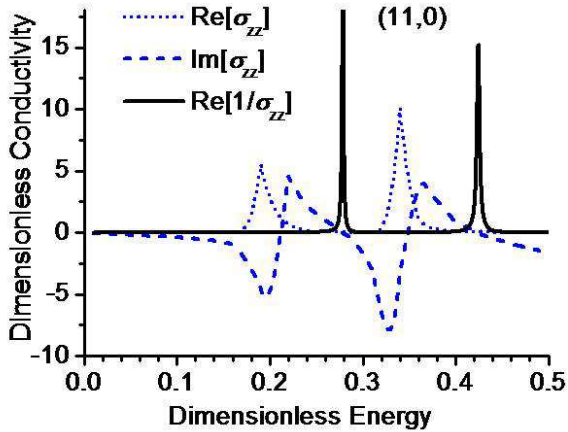}}\caption{(Color
online)~Dimensionless (in units of $e^2/2\pi\hbar$) axial surface
conductivity of the (11,0) CN per unit length. Dimensionless
energy is defined as [\emph{Energy}]/$2\gamma_0$.}\label{fig2}
\end{figure}

To compute the absorption line shapes of particular atomically
doped CNs, one needs to know their transverse and longitudinal
local photonic DOS functions
$\xi^{\perp(\parallel)}(\mathbf{r}_{A},\tilde{x}_{A})$ in
Eqs.~(\ref{Iraxfinal}) and (\ref{X}). These are determined by the
CN surface axial conductivity $\sigma_{zz}$, which is to be
calculated beforehand. Figure~\ref{fig2} shows an example of the
surface axial conductivity for the (11,0)
nanotube~\cite{Bondarev09}, calculated in the random-phase
approximation~\cite{LinShung97,EhrenreichCohen59} with the CN band
structure taken into account within the nearest-neighbor
non-orthogonal tight-binding approach~\cite{Valentin}.

\begin{figure}[b]
\epsfxsize=8.65cm\centering{\epsfbox{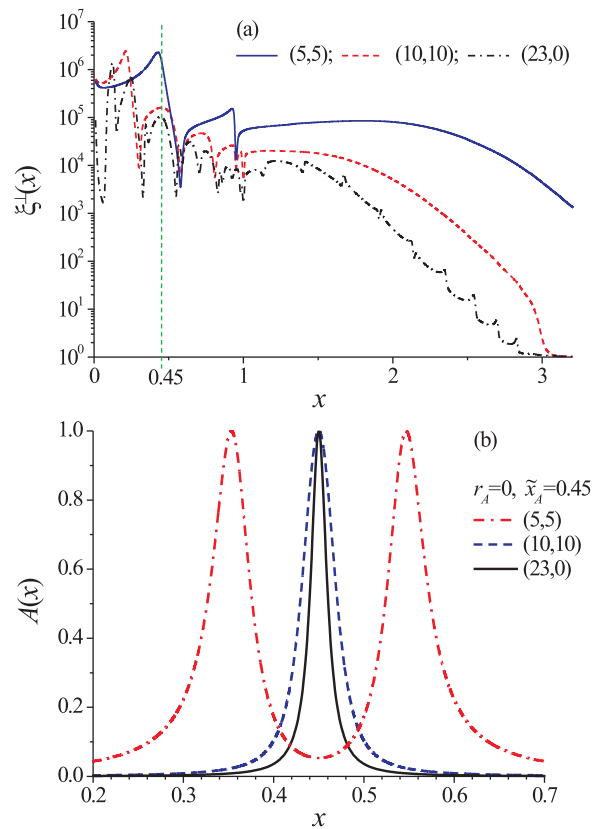}}\caption{(Color
online) Transverse local photonic DOS (a) and normalized
absorption line shapes (b) for the atom in the center of the CNs
of increasing radii. See text for notations.}\label{fig3}
\end{figure}

Figure~\ref{fig3}~(a) shows the functions $\xi^{\perp}(x)$ for the
atom in the center of the (5,5), (10,10) and (23,0) CNs
(calculated as described in
Refs.~\cite{Bondarev02,Bondarev04,Bondarev04pla,Bondarev05,Bondarev04ssc,Bondarev06trends}).
They are seen to decrease with increasing the CN radius,
representing the decrease of the atom-field coupling strength as
the atom moves away from the CN wall. To calculate the normalized
absorption curves $A(x)\!=\!\tilde{I}(x)/\tilde{I}_{peak}$ in this
case, the atomic transition frequency was fixed at
$\tilde{x}_{A}\!=\!0.45$ [vertical dashed line in
Fig.~\ref{fig3}~(a)] since this is the approximate peak position
of $\xi^{\perp}$ for all the three CNs. Figure~\ref{fig3}~(b)
shows the result. For the (5,5) CN, the absorption line is split
into two components, indicating the strong atom-field coupling
with the Rabi splitting $X\approx0.2$ corresponding to the energy
of $0.2\times2\gamma_{0}=1.08$~eV. The splitting is disappearing
with increasing the CN radius, and the absorption line shape
becomes Lorentzian with the narrower widths for the larger radii
CNs.

~

\noindent\textbf{3. Atomic States Entanglement in CNs}

~

Two identical two-level atoms (pair of atomic qubits), $A$ and
$B$, located at $\mathbf{r}_{A}\!=\!\{r_{A},0,0\}$ and
$\mathbf{r}_{B}\!=\!\{r_{B},\varphi_{B},z_{B}\}$ close to a CN,
are coupled to the \emph{common} medium-assisted vacuum-type
electromagnetic field via their transition dipole moments. Both
atomic dipole moments are, as before, assumed to be directed along
the CN axis, $\mathbf{d}_{A}=\mathbf{d}_{B}=d_{z}\textbf{e}_{z}$.
Also assumed is that the atoms are located far enough from each
other, to simplify the problem by ignoring the interatomic Coulomb
interaction. The total secondly quantized Hamiltonian of the
system of the two identical atoms and the nanotube is then written
as follows~\cite{Bondarev07} [compare with Eq.~(\ref{Htwolev})]
\begin{eqnarray}
\hat{H}=\int_{0}^{\infty}\!\!\!\!\!d\omega\,\hbar\omega\!\int\!d\mathbf{R}
\,\hat{f}^{\dag}(\mathbf{R},\omega)\hat{f}(\mathbf{R},\omega)+\!\!\!
\sum_{i=A,B}\!\!{\hbar\tilde{\omega}_{i}\over{2}}\,\hat{\sigma}_{iz}\label{Htwolev2}\\
+\sum_{i=A,B}\int_{0}^{\infty}\!\!\!\!\!d\omega\!\int\!d\mathbf{R}\;[\,
\mbox{g}^{(+)}(\mathbf{r}_{i},\mathbf{R},\omega)\,\hat{\sigma}^{\dag}_{i}\hskip1cm\nonumber\\[0.1cm]
-\,\mbox{g}^{(-)}(\mathbf{r}_{i},\mathbf{R},\omega)\,\hat{\sigma}_{i}\,]\,
\hat{f}(\mathbf{R},\omega)+\mbox{h.c.}.\hskip1.2cm\nonumber
\end{eqnarray}
The Pauli operators, $\hat{\sigma}_{i}\!=\!|L\rangle\langle
u_{i}|$, $\hat{\sigma}^{\dag}_{i}\!=\!|u_{i}\rangle\langle L|$ and
$\hat{\sigma}_{iz}\!=\!|u_{i}\rangle\langle
u_{i}|\!-\!|L\rangle\langle L|$ with $i\!=\!A\mbox{ or }B$,
describe the atomic subsystem and electric dipole transitions
between its two states, upper $|u_{i}\rangle$ with either of the
two atoms in its upper state and lower $|L\rangle$ with both atoms
in their lower states, separated by the transition frequency
$\omega_{A}$. The renormalized transition frequencies
$\tilde{\omega}_{A,B}$ are the same for both of the atoms.

For single-quantum excitations, the time-dependent wave function
of the whole system can be written as
\begin{eqnarray}
|\psi(t)\rangle=\!\!\!\sum_{i=A,B}\!\!C_{\displaystyle{u_{i}}}(t)\,
e^{-i(\tilde{\omega}_{i}-\overline{\omega})t}|u_{i}\rangle|\{0\}\rangle\hskip0.5cm\label{wfunc2}\\
+\int_{0}^{\infty}\!\!\!\!\!d\omega\!\!\int\!d\mathbf{R}\,C_{L}(\mathbf{R},\omega,t)
\,e^{-i(\omega-\overline{\omega\!}\,)t}|L\rangle|1(\mathbf{R},\omega)\rangle,\nonumber
\end{eqnarray}
where
$\overline{\omega}=\!\sum_{i=A,B}\tilde{\omega}_{i}/2\!=\!\tilde{\omega}_{A}$,
$C_{\displaystyle u_{i}}$ and $C_{L}$ are the probability
amplitudes for the upper states and the lower state of the system,
respectively. For the following it is convenient to introduce the
new variables
\[
C_{\pm}(t)=\frac{1}{\sqrt{2}}\left[C_{\displaystyle u_{A}}(t)\pm
C_{\displaystyle u_{B}}(t)\right]
\]
that are the expansion coefficients of the
wave function~(\ref{wfunc2}) in terms of the maximally entangled
two-qubit atomic states
$|\pm\rangle\!=\!(|u_{A}\rangle\pm|u_{B}\rangle)/\sqrt{2}$. In
view of Eqs.~(\ref{Htwolev2}) and (\ref{wfunc2}), the
time-dependent Schr\"{o}dinger equation yields then (dimensionless
variables --- Section 2)
\begin{equation}
\mbox{\it\.{C}}_{\pm}(\tau)=\int_{0}^{\tau}\!\!\!d\tau^{\prime}
K_{\pm}(\tau-\tau^{\prime})\,C_{\pm}(\tau^{\prime})+f_{\pm}(\tau)\,,\label{popampu2}
\end{equation}
where
\begin{eqnarray}
K_{\pm}(\tau-\tau^{\prime})=-\frac{1}{2\pi}\int_{0}^{\infty}\!\!\!dx\,\tilde{\Gamma}_{0}(x)\hskip1.25cm\label{Kpm}\\[0.1cm]
\times\;\xi^{\pm}(\mathbf{r}_{A},\mathbf{r}_{B},x)\,e^{-i(x-\tilde{x}_{A})(\tau-\tau^{\prime})},\hskip-0.5cm\nonumber
\end{eqnarray}
\begin{eqnarray}
\xi^{\pm}(\mathbf{r}_{A},\mathbf{r}_{B},x)=\frac{x^{2}_{A}}{x^{2}}\left[
\xi^{\perp}(\mathbf{r}_{A},x)\pm\xi^{\perp}(\mathbf{r}_{A},\mathbf{r}_{B},x)\right]\label{ksipm}\\[0.1cm]
+\,\xi^{\parallel}(\mathbf{r}_{A},x)\pm\xi^{\parallel}(\mathbf{r}_{A},\mathbf{r}_{B},x)\,,\hskip1.2cm\nonumber
\end{eqnarray}
\begin{equation}
f_{\pm}(\tau)=-\frac{1}{\sqrt{2}}\int_{-\Delta\tau}^{0}\!\!\!\!\!d\tau^{\prime}
K_{\pm}(\tau-\tau^{\prime})\,C_{\displaystyle{u_{A}}}(\tau^{\prime})\,.
\label{fpm}
\end{equation}
The functions $f_{\pm}(\tau)$ are only unequal to zero when the
two atoms are initially in their ground states, with the initial
excitation residing in the nanotube. Equation~(\ref{fpm}) assumes
that this is realized by selecting the time origin to be right
after the time interval $\Delta\tau$, that is necessary for the
(excited) atom $A$ to decay completely into the nanotube photonic
modes, has elapsed. The two-particle local photonic DOS functions
\[
\xi^{\perp(\parallel)}(\mathbf{r}_{A},\mathbf{r}_{B},x)=
\frac{\mbox{Im}^{\perp(\parallel)}G_{zz}^{\,\perp(\parallel)}
(\mathbf{r}_{A},\mathbf{r}_{B},x)}{\mbox{Im}G_{zz}^{0}(x)}
\]
are the generalizations of the DOS functions
$\xi^{\perp(\parallel)}(\mathbf{r}_{A},x)$ of Section~2 (see
Refs.~\cite{Bondarev04,Bondarev05,Bondarev06trends,Bondarev07} for
more details).

The entanglement of the two quantum bits occurs when the two-qubit
wave function cannot be represented as a~product of the two
one-qubit states in any basis. To determine this quantity, a
recipe based on the "spin flip" transformation (see
Ref.~\cite{Wooters} for details) and valid for an arbitrary number
of the qubits is used. First, the reduced density matrix is
defined
\[
\hat{\rho}_{AB}(\tau)=|\psi_{AB}(\tau)\rangle\langle\psi_{AB}(\tau)|
=\mbox{Tr}_{field}|\psi(\tau)\rangle\langle\psi(\tau)|\,,
\]
that describes the bipartite atomic subsystem in terms of the wave
function~(\ref{wfunc2}) of the whole system. Next, the
"concurrence"
\[
\mbox{C}(\psi_{AB})=|\langle\psi_{AB}|\tilde{\psi}_{AB}\rangle|
\]
is introduced, where $|\tilde{\psi}_{AB}\rangle\!=
\!\hat{\sigma}_{y}^{A}\hat{\sigma}_{y}^{B}|\psi_{AB}^{\ast}\rangle$
with $\hat{\sigma}_{y}^{A(B)}$ being the Pauli matrix that
represents the "spin flip" transformation in the atom $A(B)$
single-qubit space.~This, after some algebra, becomes
\begin{equation}
\mbox{C}[\psi_{AB}(\tau)]=|C_{+}^{2}(\tau)-C_{-}^{2}(\tau)|
\label{conc}
\end{equation}
with $C_{\pm}(\tau)$ given by the integral
equation~(\ref{popampu2}). Finally, the degree of the entanglement
of the two-qubit atomic state $|\psi_{AB}\rangle$ ("entanglement
of formation"~\cite{Wooters}) is given by
\begin{equation}
\mbox{E}[\psi_{AB}(\tau)]=h\!\left\{1+\frac{1}{2}\sqrt{1-\mbox{C}[\psi_{AB}(\tau)]^{2}}\right\},
\label{entangl}
\end{equation}
where $h(y)\!=\!-y\log_{2}y-(1-y)\log_{2}(1-y)$.

The entanglement $\mbox{E}[\psi_{AB}(\tau)]$ is maximal when the
coefficients $C_{+}(\tau)$ and $C_{-}(\tau)$ are maximally
different from each other. For this to occur, the functions
$\xi^{\pm}(\mathbf{r}_{A},\mathbf{r}_{B},x)$ in
Eqs.~(\ref{popampu2})--(\ref{fpm}) should be different in their
values. These are determined by
$\xi^{\,\perp(\parallel)}(\mathbf{r}_{A},\mathbf{r}_{B},x)$ whose
frequency behavior is determined by the CN surface axial
conductivity $\sigma_{zz}$ (Fig.~\ref{fig2}). The functions
$\xi^{\,\perp(\parallel)}(\mathbf{r}_{A},\mathbf{r}_{B},x)$ are
computed as described in Ref.~\cite{Bondarev07}. Then, the
integral equation (\ref{popampu2}) is solved numerically to obtain
$C_{\pm}(\tau)$, followed by the "concurence" (\ref{conc}) and the
entanglement of formation (\ref{entangl}).

\begin{figure}[t]
\epsfxsize=8.65cm\centering{\epsfbox{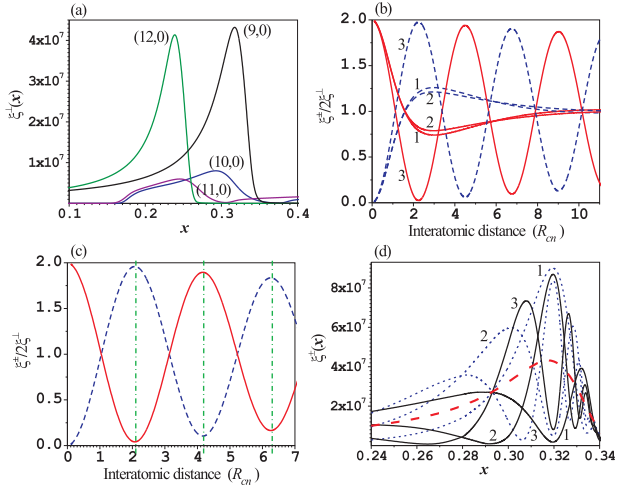}} \caption{(Color
online) (a)~Transverse local photonic DOS functions
$\xi^{\perp}(x)$ for the two-level atom in the centers of the four
'zigzag' nanotubes. (b)~Normalized two-particle local photonic DOS
functions $\xi^{+}$ (solid lines) and $\xi^{-}$ (dashed lines)
taken at the peak frequencies of $\xi^{\perp}(x)$ [see (a)], as
functions of the distances between the two atoms on the axes of
the (10,0) (lines~1; $x\!=\!0.29$), (11,0) (lines~2; $x\!=\!0.25$)
and (12,0) (lines~3; $x\!=\!0.24$) CNs. (c)~Same as in (b) for the
two atoms on the axis of the (9,0) CN ($x\!=\!0.32$).
(d)~Two-particle DOS functions $\xi^{+}(x)$ (solid lines) and
$\xi^{-}(x)$ (dotted lines) for the two atoms located in the
center of the (9,0) CN and separated from each other by the
distances of $2.1R_{cn}\!\approx\!7.4\,$\AA\space (lines~1),
$4.2R_{cn}\!\approx\!14.8\,$\AA\space (lines~2) and
$6.3R_{cn}\!\approx\!22.2\,$\AA\space (lines~3) [shown by the
vertical lines in (c)]; the dashed line shows $\xi^{\perp}(x)$ for
the atom in the center of the (9,0) CN.} \label{fig4}
\end{figure}

Shown in Fig.~\ref{fig4}~(a) is a typical frequency behavior of
the one-particle transverse photonic DOS
$\xi^{\perp}(\mathbf{r}_{A}\!=\!0,x)$ in the infrared and visible
frequency range $x<0.4$ for the atom inside 'zigzag' CNs of
increasing radii. The DOS resonances are seen to be much sharper
for metallic CNs ($m\!=\!3q,~q\!=\!1,2,...$) than for
semiconducting ones ($m\!\ne\!3q$) in agreement with the fact that
this frequency range is dominated by the classical Drude-type
conductivity which is larger in metallic CNs~\cite{Bondarev02}.
Figure~\ref{fig4}~(b) shows the normalized two-particle local
photonic DOS functions $\xi^{\pm}$ taken at the resonance
frequencies of the respective $\xi^{\perp}$'s
[Fig.~\ref{fig4}~(a)], as functions of the distance between the
two atoms on the axes of the (10,0), (11,0) and (12,0) CNs. The
values of $\xi^{+}$ and $\xi^{-}$ are seen to be substantially
different from each other before they reach their limit values of
$\xi^{+}=\xi^{-}=\xi^{\perp}+\xi^{\parallel}\approx2\xi^{\perp}$
at sufficiently large interatomic separations. For metallic CNs
the functions $\xi^{+}$ and $\xi^{-}$ exhibit resonator-like
behavior, i.e. they vary periodically in antiphase almost without
damping with increasing interatomic separation. In
Fig.~\ref{fig4}~(c) is shown the dependence of
$\xi^{\pm}(x\!=\!0.32)$ [peak position of $\xi^{\perp}$ in
Fig.~\ref{fig4}~(a)] on the interatomic separation for the atoms
in the center of the metallic (9,0) CN. As the separation
increases, $\xi^{+}$ and $\xi^{-}$ may differ greatly one from
another in a periodic way. The maximal difference is
$|\xi^{+}-\xi^{-}|\approx2(\xi^{\perp}+\xi^{\parallel})
\approx4\xi^{\perp}\!\sim\!10^{8}$, making the mixing coefficients
$C_{\pm}(\tau)$ different and thus resulting in a substantial
degree of the entanglement of the two spatially separated atomic
qubits.

\begin{figure}[t]
\epsfxsize=8.65cm\centering{\epsfbox{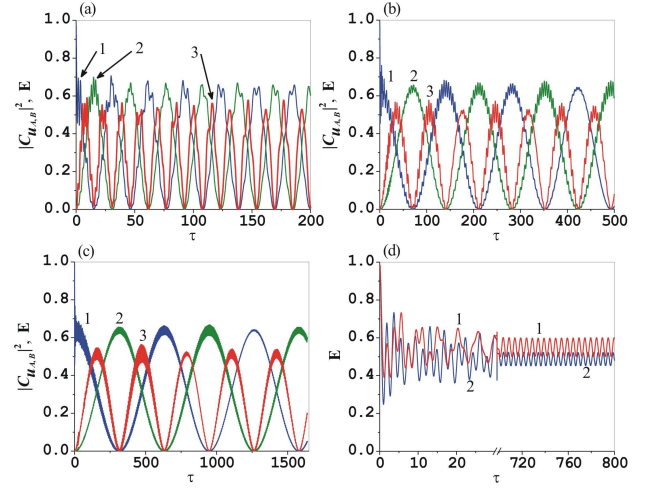}} \caption{(Color
online) (a),(b),(c)~Upper decay of the initially excited atom~$A$
(lines~1) and initially unexcited atom~$B$ (lines~2), and the
entanglement (lines~3), as functions of dimensionless time for the
two atoms located in the center of the (9,0) CN and separated from
each other by the distances of $2.1R_{cn}\!\approx\!7.4\,$\AA,
$4.2R_{cn}\!\approx\!14.8\,$\AA\space and
$6.3R_{cn}\!\approx\!22.2\,$\AA, respectively [the situation shown
in terms of local photonic DOS in Figs.~\ref{fig4}(c),(d)].
(d)~Short-time and long-time entanglement evolution of the
initially fully entangled atoms separated by $7.4\,$\AA; lines 1
and 2 are for $C_{+}(0)\!=\!1, C_{-}(0)\!=\!0$ and $C_{+}(0)=0,
C_{-}(0)=1$, respectively.} \label{fig5}
\end{figure}

The ensuing spontaneous decay dynamics and atomic entanglement are
presented in Fig.~\ref{fig5}~(a),(b),(c) for the atoms separated
by the three different distances [shown by the vertical dashed
lines in Fig.~\ref{fig4}~(c)] in the center of the (9,0) CN. Atom
$A$ is assumed to be initially excited while atom $B$ is in its
ground state. The entanglement is seen to reach the amount of 0.5
and to vary with time periodically without damping at least for
the (reasonably long) times we restricted ourselves in our
computations. As the interatomic separation increases, so the
period of the entanglement oscillations does while no change
occurs in the maximal entanglement. Figure~\ref{fig5}~(d) shows
the short-time and long-time entanglement evolution when both of
the atoms are initially maximally entangled [$C_{\pm}(0)\!=\!1$
while $C_{\mp}(0)\!=\!0$] and separated by the distance of
$7.4$~\AA. The entanglement is slightly larger in the case where
$C_{+}(0)\!=\!1,C_{-}(0)\!=\!0$ and no damping occurs as
before.~Note that the atoms can be separated by longer distances
with roughly the same entanglement due to the $\xi^{\pm}$
periodicity with interatomic distance.

~

\noindent\textbf{4. Exciton-Surface-Plasmon Coupling in Pristine
Semiconducting CNs}

~

The total Hamiltonian of an exciton interacting with vacuum-type
electromagnetic fluctuations (no external electromagnetic field is
assumed to be applied) on the surface of a pristine single-walled
semiconducting carbon nanotube is of the form~\cite{Bondarev08}
\begin{equation}
\hat{H}=\hat{H}_F+\hat{H}_{ex}+\hat{H}_{int}\,,\label{Htot}
\end{equation}
where the three terms represent the free field, the free exciton,
and their interaction, respectively. Here, the second quantized
electromagnetic field Hamiltonian is
\begin{equation}
\hat{H}_F=\sum_\mathbf{n}\int_0^\infty\!d\omega\,\hbar\omega\,
\hat{f}^\dag(\mathbf{n},\omega)\hat{f}(\mathbf{n},\omega)
\label{Hf}
\end{equation}
with the scalar bosonic field operators
$\hat{f}^\dag(\mathbf{n},\omega)$ and $\hat{f}(\mathbf{n},\omega)$
creating and annihilating, respectively, the (vacuum-type)
\emph{medium-assisted} surface electromagnetic excitation of
frequency $\omega$ at an arbitrary point
$\mathbf{n}\!=\!\mathbf{R}_n\!=\!\{R_{CN},\varphi_n,z_n\}$
associated with a carbon atom (representing a lattice site) on the
surface of the CN of radius $R_{CN}$. The summation is made over
all the carbon atoms, and may be replaced by the integration over
the entire nanotube surface according to the rule
\[
\sum_\mathbf{n}\!\ldots=\frac{1}{S_0}\!\int\!d\mathbf{R}_n\!\ldots=
\frac{1}{S_0}\!\int_0^{2\pi}\!d\varphi_nR_{CN}\int_{-\infty}^\infty\!dz_n\!\ldots\,,
\]
where $S_0\!=\!(3\sqrt{3}/4)b^2$ is the area of an elementary
equilateral triangle selected around each carbon atom in a way to
cover the entire surface of the nanotube, with
$b\!=\!1.42$~\AA\space being the carbon-carbon interatomic
distance.

\begin{figure}[t]
\epsfxsize=8.65cm\centering{\epsfbox{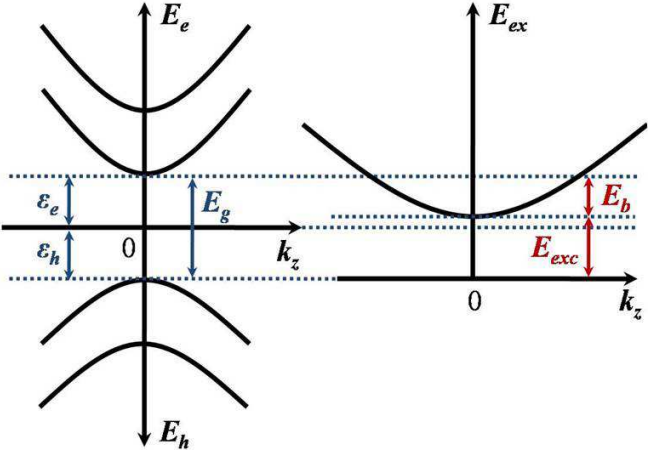}}\caption{(Color
online)~Schematic of the transversely quantized azimuthal
electron-hole subbands (\emph{left}), and the first-interband
ground-internal-state exciton energy (\emph{right}) in a
small-diameter semiconducting carbon nanotube. See text for
notations.}\label{fig6}
\end{figure}

The second quantized Hamiltonian of the free exciton (see, e.g.,
Ref.~\cite{Haken}) is of the following form
\begin{equation}
\hat{H}_{ex}=\sum_{\mathbf{n},\mathbf{m},f}\!E_f(\mathbf{n})
B^\dag_{\mathbf{n}+\mathbf{m},f}B_{\mathbf{m},f}\!=
\!\sum_{\mathbf{k},f}\!E_f(\mathbf{k})B^\dag_{\mathbf{k},f}B_{\mathbf{k},f},
\label{Hex}
\end{equation}
where the operators $B^\dag_{\mathbf{n},f}$ and $B_{\mathbf{n},f}$
create and annihilate, respectively, an exciton with energy
$E_f(\mathbf{n})$ at the lattice site $\mathbf{n}$ of the CN
surface.~The index $f\,(\ne\!0)$ refers to the internal degrees of
freedom of the exciton. Alternatively,
$B^\dag_{\mathbf{k},f}\!=\!\sum_{\mathbf{n}}\!B^\dag_{\mathbf{n},f}
e^{i\mathbf{k}\cdot\mathbf{n}}/\sqrt{N}$ creates ($N$ is the
number of the lattice sites) and
$B_{\mathbf{k},f}\!=\!(B^\dag_{\mathbf{k},f})^\dag$ annihilates
the $f$-internal-state exciton with the quasi-momentum
$\mathbf{k}\!=\!\{k_{\varphi},k_{z}\}$, where $k_{\varphi}$ is
quantized due to the transverse confinement effect and $k_{z}$ is
continuous. The exciton total energy is given by
\begin{equation}
E_f(\mathbf{k})=E_{exc}^{(f)}(k_{\varphi})+\frac{\hbar^2k_z^2}{2M_{ex}}
\label{Ef}
\end{equation}
with the first term representing the excitation energy,
$E_{exc}^{(f)}(k_{\varphi})=E_g(k_{\varphi})+E_b^{(f)}$, of the
$f$-internal-state exciton with the (negative) binding energy
$E_b^{(f)}$, created via the interband transition with the band
gap
$E_g(k_{\varphi})\!=\!\varepsilon_e(k_{\varphi})+\varepsilon_h(k_{\varphi})$,
where $\varepsilon_{e,h}$ are transversely quantized azimuthal
electron-hole subbands (see the schematic in Fig.~\ref{fig6}). The
second term represents the kinetic energy of the translational
longitudinal movement of the exciton with the effective mass
$M_{ex}=m_e+m_h$, where $m_e$ and $m_h$ are the electron and hole
effective masses, respectively. The two equivalent free-exciton
Hamiltonian representations are related to one another via the
obvious orthogonality relationships
$\sum_\mathbf{n}e^{-i(\mathbf{k}-\mathbf{k}^\prime)\cdot\mathbf{n}}/N\!=
\!\delta_{\mathbf{k}\mathbf{k}^\prime}$ and
$\sum_\mathbf{k}e^{-i(\mathbf{n}-\mathbf{m})\cdot\mathbf{\mathbf{k}}}/N\!=
\!\delta_{\mathbf{n}\mathbf{m}}$ with the $\mathbf{k}$-summation
running over the first Brillouin zone of the nanotube.~The bosonic
field operators in Eq.~(\ref{Hf}) are transformed to the
$\mathbf{k}$-representation in the same way.

The most general (non-relativistic, electric dipole)
exciton-electromagnetic-field interaction on the nanotube surface
can be written in the
form~\cite{Bondarev08,Bondarev05,Bondarev06trends}
\begin{equation}
\hat{H}_{int}=\hat{H}_{int}^{(1)}+\hat{H}_{int}^{(2)}\label{Hint}
\end{equation}\vspace{-0.5cm}
\[
=-\frac{e}{m_{e}c}\sum_\mathbf{n}\hat{\mathbf{A}}(\mathbf{n})\!\cdot\!
\left[\hat{\mathbf{p}}_\mathbf{n}-\frac{e}{2c}\hat{\mathbf{A}}(\mathbf{n})\right]
+\sum_\mathbf{n}\hat{\mathbf{d}}_\mathbf{n}\!\cdot\mathbf{\nabla}_{\!\mathbf{n}}
\hat{\varphi}(\mathbf{n})\,,
\]
where
$\hat{\mathbf{p}}_\mathbf{n}\!=\!\sum_f\langle0|\hat{\mathbf{p}}_\mathbf{n}|f\rangle
B_{\mathbf{n},f}+h.c.\,$ is the total electron momentum operator
at the lattice site~$\mathbf{n}$ under the optical dipole
transition resulting in the exciton formation at the same site,
$\hat{\mathbf{d}}_\mathbf{n}\!=\!\sum_f\langle0|\hat{\mathbf{d}}_\mathbf{n}|f\rangle
B_{\mathbf{n},f}+h.c.$ is the corresponding transition dipole
moment operator [related to $\hat{\mathbf{p}}_\mathbf{n}$ via the
equation
$\langle0|\hat{\mathbf{p}}_\mathbf{n}|f\rangle\!=\!im_{e}E_{f}(\mathbf{n})
\langle0|\hat{\mathbf{d}}_\mathbf{n}|f\rangle/\hbar e$], $c$ and
$e$ are the speed of light and the electron charge,
respectively.~The vector potential operator
$\hat{\mathbf{A}}(\mathbf{n})$ (the Coulomb gauge is assumed) and
the scalar potential operator $\hat{\varphi}(\mathbf{n})$
represent, respectively, the nanotube's transversely polarized
surface electromagnetic modes and longitudinally polarized surface
electromagnetic modes which the exciton interacts with,
\begin{eqnarray}
\hat{\mathbf{A}}(\mathbf{n})=\int_0^\infty\!\!\!\!\!d\omega\frac{c}{i\omega}\,
\hat{\underline{\mathbf{E}}}^\perp(\mathbf{n},\omega)+h.c.=
\sum_{\mathbf{m}}\int_0^\infty\!\!\!d\omega\hskip1cm\nonumber\\
\times\frac{4}{c}\sqrt{\pi\hbar\omega\mbox{Re}\sigma_{zz}(R_{CN},\omega)}\;
^{\perp}G_{zz}(\mathbf{n},\mathbf{m},\omega)\hat{f}(\mathbf{m},\omega)\!+\!h.c.,\nonumber\\[0.2cm]
-\mathbf{\nabla}_{\!\mathbf{n}}\hat{\varphi}(\mathbf{n})=\int_0^\infty\!\!\!\!\!d\omega\,
\hat{\underline{\mathbf{E}}}^\parallel(\mathbf{n},\omega)+h.c.=
\sum_{\mathbf{m}}\int_0^\infty\!\!\!d\omega\hskip1cm\nonumber\\
\times\frac{4i\omega}{c^2}\sqrt{\pi\hbar\omega\mbox{Re}\sigma_{zz}(R_{CN},\omega)}\;
^{\parallel}G_{zz}(\mathbf{n},\mathbf{m},\omega)\hat{f}(\mathbf{m},\omega)\!+\!h.c.,\nonumber
\end{eqnarray}
with the total electric field operator given by
\[
\hat{\mathbf{E}}(\mathbf{n})\!=\!\int_0^\infty\!\!\!d\omega
\left[\,\hat{\underline{\mathbf{E}}}^\perp(\mathbf{n},\omega)+
\hat{\underline{\mathbf{E}}}^\parallel(\mathbf{n},\omega)\right]+h.c.
\]
and $^{\perp(\parallel)}G_{zz}(\mathbf{n},\mathbf{m},\omega)$
representing the $zz$-component of the transverse (longitudinal)
Green tensor of the electromagnetic subsystem (the nanotube). This
Green tensor is different from that used for atomically doped CNs
in Sections~2 and 3 above since now there is no additional atomic
source in the problem. The Green tensor is derived by expanding
the solution of the Green equation in cylindrical coordinates and
determining the Wronskian normalization constant from the
appropriately chosen boundary conditions on the CN
surface~\cite{Jackson,Bondarev04,Bondarev06trends}.

Equations~(\ref{Htot})--(\ref{Hint}) form the complete set of
equations describing the exciton-photon coupled system on the CN
surface in terms of the electromagnetic field Green tensor and the
CN surface axial conductivity.

It is important to realize that the transversely polarized surface
electromagnetic mode contribution to the interaction Hamiltonian
from Eq.~(\ref{Hint}) (first term) is negligible compared to the
longitudinally polarized surface electromagnetic mode contribution
(second term).~The point is that, because of the nanotube
quasi-one-dimensionality, the exciton quasi-momentum vector and
all the relevant vectorial matrix elements of the momentum and
dipole moment operators are directed predominantly along the CN
axis (the longitudinal exciton). This prevents the exciton from
the electric dipole coupling to transversely polarized surface
electromagnetic modes as they propagate predominantly along the CN
axis with their electric vectors orthogonal to the propagation
direction.~The longitudinally polarized surface electromagnetic
modes are generated by the electronic Coulomb potential (see,
e.g., Ref.~\cite{Landau}), and therefore represent the CN surface
plasmon excitations.~These have their electric vectors directed
along the propagation direction. They do couple to the
longitudinal excitons on the CN surface. Such modes were observed
in Ref.~\cite{Pichler98}. They occur in CNs both at high energies
(well-known $\pi$-plasmon~at $\sim\!6$~eV) and at comparatively
low energies of $\sim\!0.5\!-\!2$~eV.~The latter ones are related
to transversely quantized interband (inter-van Hove) electronic
transitions. These weakly-dispersive
modes~\cite{Pichler98,Kempa02} are similar to the intersubband
plasmons in quantum wells~\cite{Kempa89}. They occur in the same
energy range of $\sim\!1$~eV where the exciton excitation energies
are located in small-diameter ($\lesssim\!1$~nm) semiconducting
CNs~\cite{Spataru05,Ma}.

The Hamiltonian~(\ref{Htot})--(\ref{Hint}) can be diagonalized
exactly using Bogoliubov's canonical transformation technique
(see, e.g., Ref.~\cite{Davydov}), to yield~\cite{Bondarev08}
\begin{equation}
\hat{H}=\sum_{\mathbf{k},\,\mu=1,2}\hbar\omega_\mu(\mathbf{k})\,
\hat{\xi}^\dag_\mu(\mathbf{k})\hat{\xi}_\mu(\mathbf{k})+E_0\,,
\label{Htotdiag}
\end{equation}
where the new operator

\begin{eqnarray}
\hat{\xi}_\mu(\mathbf{k})=\sum_{f}\left[\,u_\mu^\ast(\mathbf{k},\omega_f)B_{\mathbf{k},f}
-v_\mu(\mathbf{k},\omega_f)B_{-\mathbf{k},f}^\dag\right]\nonumber\\
+\int_0^\infty\!\!\!d\omega\left[\,u_\mu(\mathbf{k},\omega)\hat{f}(\mathbf{k},\omega)-
v_\mu^\ast(\mathbf{k},\omega)\hat{f}^\dag(-\mathbf{k},\omega)\right]\nonumber
\end{eqnarray}
annihilates and
$\hat{\xi}^\dag_\mu(\mathbf{k})\!=\![\hat{\xi}_\mu(\mathbf{k})]^\dag$
creates the coupled exciton-plasmon excitation of branch
$\mu\,(=\!1,2)$, $u_\mu$ and $v_\mu$ are the (appropriately
chosen) canonical transformation coefficients,
$\omega_f\!=\!E_f/\hbar$. The "vacuum" energy $E_0$ represents the
state with no exciton-plasmons excited in the system, and
$\hbar\omega_\mu(\mathbf{k})$ is the exciton-plasmon energy given
by the solution of the following (dimensionless) dispersion
relation~\cite{Bondarev08,Bondarev09}
\begin{equation}
x_\mu^2-\varepsilon_f^2-\frac{\varepsilon_f}{2\pi}\int_0^\infty\!\!\!\!\!dx\,
x\,\tilde{\Gamma}_0^f(x)\frac{\rho(x)}{x_\mu^2-x^2}=0\,.\label{dispeq}
\end{equation}
Here,
\[
x_\mu=\frac{\hbar\omega_\mu(\mathbf{k})}{2\gamma_0}\,,~~~
\varepsilon_f=\frac{E_f(\mathbf{k})}{2\gamma_0}\,,
\]
the function
\[
\tilde{\Gamma}_0^f(x)=\frac{4|d^f_z|^2x^3}{3\hbar c^3}
\left(\frac{2\gamma_0}{\hbar}\right)^{\!2}
\]
represents the (dimensionless) exciton spontaneous decay rate,
where
$d^f_z\!=\!\sum_{\mathbf{n}}\langle0|(\hat{\mathbf{d}}_\mathbf{n})_z|f\rangle$
is the longitudinal exciton transition dipole moment matrix
element, and the function
\begin{equation}
\rho(x)=\frac{3S_0}{16\pi\alpha
R_{CN}^2}\;\mbox{Re}\frac{1}{\bar\sigma_{zz}(x)}\label{plDOS}
\end{equation}
stands for the surface plasmon DOS responsible for the exciton
decay rate variation due to its coupling to the surface plasmon
modes.~In Eq.~(\ref{plDOS}), $\alpha\!=\!e^2/\hbar c\!=\!1/137$ is
the fine-structure constant and
$\bar\sigma_{zz}\!=\!2\pi\hbar\sigma_{zz}/e^2$ is the
dimensionless CN surface axial conductivity per unit length.

\begin{figure}[t]
\epsfxsize=8.65cm\centering{\epsfbox{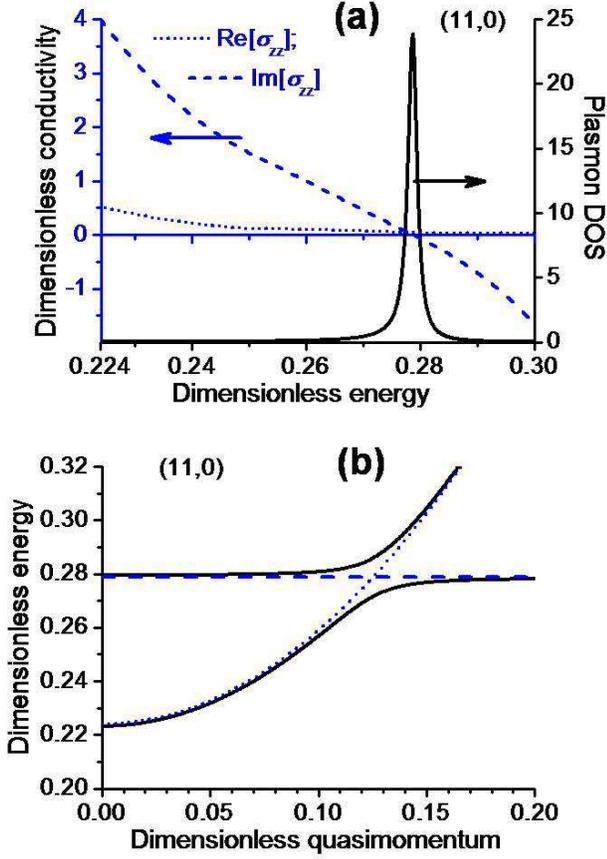}}\caption{(Color
online) Fragments of the real/imaginary conductivity and the
surface plasmon DOS~(a), and the lowest bright exciton
dispersion~(b) when coupled to plasmons in the (11,0) CN.
Dimensionless energy is defined as [\emph{Energy}]/$2\gamma_0$.
See text for dimensionless quasi-momentum.}\label{fig7}
\end{figure}

Note that the conductivity factor in Eq.~(\ref{plDOS}) equals
\[
\mbox{Re}\frac{1}{\bar\sigma_{zz}(x)}=-\frac{4\alpha
c}{R_{CN}}\left(\frac{\hbar}{2\gamma_0x}\right)\mbox{Im}\frac{1}{\epsilon_{zz}(x)-1}
\]
in view of the equation
\[
\sigma_{zz}=-\frac{i\omega(\epsilon_{zz}\!-\!1)}{4\pi S\rho_{T}}
\]
representing the Drude relation for CNs, where
$\epsilon_{zz}$ is the longitudinal dielectric function, $S$ and
$\rho_{T}$ are the surface area of the tubule and the number of
tubules per unit volume,
respectively~\cite{Bondarev04,Bondarev05,Bondarev06trends,Tasaki}.~This
relates very closely the surface plasmon DOS
function~(\ref{plDOS}) to the loss function
$-\mbox{Im}(1/\epsilon)$ measured in Electron Energy Loss
Spectroscopy (EELS) experiments to determine the properties of
collective electronic excitations in solids~\cite{Pichler98}.

Figure~\ref{fig2} above shows the low-energy behavior of the
functions $\bar\sigma_{zz}(x)$ and
$\mbox{Re}[1/\bar\sigma_{zz}(x)]$ for the (11,0) nanotube taken as
an example ($R_{CN}=0.43$~nm). The function
$\mbox{Re}[1/\bar\sigma_{zz}(x)]$ is~only non-zero when the two
conditions, $\mbox{Im}[\bar\sigma_{zz}(x)]=0$ and
$\mbox{Re}[\bar\sigma_{zz}(x)]\rightarrow0$, are fulfilled
simultaneously~\cite{Kempa02,Kempa89,LinShung97}. These result in
the peak structure of $\mbox{Re}(1/\bar\sigma_{zz})$ and of the
surface plasmon DOS function (\ref{plDOS}), respectively. The
plasmon DOS peaks represent the surface plasmon modes associated
with the transversely quantized interband electronic transitions
in CNs~\cite{Kempa02} (transitions in between the azimuthal
electron-hole subbands in the left panel of Fig.~\ref{fig6}).

Figure~\ref{fig7}~(a) shows the lowest-energy plasmon DOS
resonance of the (11,0) CN~\cite{Bondarev09} as given by
Eq.~(\ref{plDOS}). The corresponding fragments of the nanotube's
real and imaginary conductivities are also shown there for
comparison. The lower dimensionless energy limit is set up to be
equal to the lowest bright exciton excitation energy
[$E_{exc}=1.21$~eV ($x=0.224$) for the (11,0) CN as reported in
Ref.\cite{Spataru05} by directly solving the Bethe-Salpeter
equation]. The peak of the DOS function is seen to exactly
coincide in energy with the zero of
$\mbox{Im}[\bar\sigma_{zz}(x)]$ \{or the zero of
$\mbox{Re}[\epsilon_{zz}(x)]$\}, clearly indicating the plasmonic
nature of the CN surface excitations under
consideration~\cite{Kempa02,Kempa}. Also seen in Fig.~\ref{fig7}
that the interband plasmon excitations occur in CNs slightly above
the first bright exciton excitation energy~\cite{Ando}. This is a
unique feature of the complex dielectric response function
--- the consequence of the general Kramers-Kr\"{o}nig
relation~\cite{VogelWelsch}.

The sharp peak structure of the surface plasmon DOS $\rho(x)$ may
be used to solve the dispersion equation~(\ref{dispeq})
analytically. Using the Lorentzian approximation
\[
\rho(x)\approx\frac{\rho(x_p)\Delta x_{p}^2} {(x-x_{p})^2+\Delta
x_{p}^2}
\]
with $x_{p}$ and $\Delta x_{p}$ being, respectively, the position
and the half-width-at-half-maximum of the plasmon resonance
closest to the lowest bright exciton excitation energy, the
integral in Eq.~(\ref{dispeq}) simplifies to the form
\begin{eqnarray}
\frac{2}{\pi}\int_0^\infty\!\!\!\!\!dx\,\frac{x\,\tilde{\Gamma}_0^f(x)\rho(x)}{x_\mu^2-x^2}
\approx\frac{F(x_p)\Delta x_{p}^2}{x_\mu^2-x_p^2}\!
\int_0^\infty\!\!\!\!\!\frac{dx}{(x-x_{p})^2+\Delta\,\!x_{p}^2}\nonumber\\[0.2cm]
=\frac{F(x_p)\Delta\,\!x_{p}}{x_\mu^2-x_p^2}\left[\arctan\!\left(\frac{x_p}{\Delta\,\!x_p}\right)+
\frac{\pi}{2}\right]\hskip1.4cm\nonumber
\end{eqnarray}
where $F(x_p)=2x_p\tilde{\Gamma}_0^f(x_p)\rho(x_p)/\pi$. This is
valid for all $x_\mu$ in Eq.~(\ref{dispeq}) apart from those
located in the narrow interval $(x_p-\Delta x_p,x_p+\Delta x_p)$
in the vicinity of the plasmon resonance, provided that the
resonance is sharp enough. Then, the dispersion equation becomes
the biquadratic equation for $x_\mu$ with the following two
positive solutions of interest (the dispersion curves)
\begin{equation}
x_{1,2}=\sqrt{\frac{\varepsilon_f^2+x_{p}^2}{2}\pm\frac{1}{2}
\sqrt{(\varepsilon_f^2\!-x_{p}^2)^2+F_{\!p}\,\varepsilon_f}}\,.
\label{dispsol}
\end{equation}
Here, $F_{\!p}=4F(x_p)\Delta\,\!x_p(\pi-\Delta\,\! x_{p}/x_{p})$
(the $\arctan$-function is expanded into series to linear terms in
$\Delta x_p/x_p\ll1$).

The two dispersion curves of Eq.~(\ref{dispsol}) are shown in
Fig.~\ref{fig7}~(b) for the (11,0) nanotube~\cite{Bondarev09} as
functions of the dimensionless longitudinal quasi-momentum.~In
these calculations, the dipole interband transition matrix element
in $\bar\Gamma_0^f(x_p)$ was estimated from the equation
$|d^f_z|^2=3\hbar\lambda^3/4\tau_{ex}^{rad}$ (according to
Hanamura's general theory of the exciton radiative decay in
spatially confined systems~\cite{Hanamura}), where
$\tau_{ex}^{rad}$ is the exciton intrinsic radiative lifetime, and
$\lambda=2\pi c\hbar/E$ with $E$ being the exciton total energy
given by Eq.~(\ref{Ef}).~For zigzag-type nanotubes, the first
Brillouin zone of the longitudinal quasi-momentum is given by
$-2\pi\hbar/3b\le\hbar k_z\le2\pi\hbar/3b$~\cite{Dresselhaus,Dai}.
The total energy of the ground-internal-state exciton can then be
written as $E=E_{exc}+(2\pi\hbar/3b)^2t^2/2M_{ex}$ with $-1\le
t\le1$ representing the dimensionless longitudinal quasi-momentum.
The lowest bright exciton parameters $E_{exc}=1.21$~eV,
$\tau_{ex}^{rad}=14.3$~ps, and $M_{ex}=0.44m_0$ ($m_0$ is the
free-electron mass) were used for the (11,0) CN as reported in
Ref.\cite{Spataru05} from the numerical solution of the
Bethe-Salpeter equation.

The dispersion curves in Fig.~\ref{fig7}~(b) demonstrate a clear
anticrossing behavior with the (Rabi) energy splitting
$\sim\!0.1$~eV. This indicates the formation of the strongly
coupled surface plasmon-exciton excitations. Note that here the
strong exciton-plasmon interaction is supported by an individual
quasi-1D nanostructure, a single-walled (small-diameter)
semiconducting carbon nanotube, as opposed to the artificially
fabricated metal-semiconductor nanostructures studied
previously~\cite{Bellessa,Govorov,Fedutik} where the metallic
component normally carries the plasmon and the semiconducting one
carries the exciton. It is also important that the effect comes
not only from the height but also from the width of the plasmon
resonance as is seen from the definition of the $F_p$ factor in
Eq.~(\ref{dispsol}). In other words, as long as the plasmon
resonance is sharp enough (which is always the case for interband
plasmons), the effect is determined by the area under the plasmon
peak in the DOS function~(\ref{plDOS}) rather then by the peak
height as one would expect.

The theory of Section~2 can be applied to obtain the optical
absorption lineshape (optical response) for the surface
exciton-plasmons in CNs. A slight difference now is that the
exciton-phonon scattering should also be taken into account as a
concurrent process to scatter the exciton in addition to the
exciton-plasmon scattering. In the relaxation time approximation
for the exciton-phonon scattering, the (dimensionless) exciton
absorption lineshape function $\tilde{I}(x)$ in the vicinity of
the plasmon resonance is then of the form~\cite{Bondarev09}
\begin{equation}
\tilde{I}(x)\!=\!\frac{\tilde{I}_{0}(\varepsilon_f)\left[(x-\varepsilon_f)^{2}+\Delta\,\!x_p^2\,\right]}
{[(x\!-\!\varepsilon_f)^{2}\!-\!X_f^{2}/4]^{2}+(x\!-\!\varepsilon_f)^{2}(\Delta\,\!x_p^2\!+\!\Delta\varepsilon_f^2)}
\label{Ixfin}
\end{equation}
with
\[
\tilde{I}_{0}(\varepsilon_f)=\frac{1}{2\pi}\tilde{\Gamma}_0^f(\varepsilon_f)\rho(\varepsilon_f)\,,\;\;\;
X_f=\sqrt{4\pi\Delta x_p\,\tilde{I}_{0}(\varepsilon_f)}\,,
\]
and $\Delta\varepsilon_f\!=\!\hbar/2\gamma_0\tau_{ph}$ being the
exciton energy broadening due to the exciton-phonon scattering
with the relaxation time $\tau_{ph}$.

\begin{figure}[t]
\epsfxsize=8.65cm\centering{\epsfbox{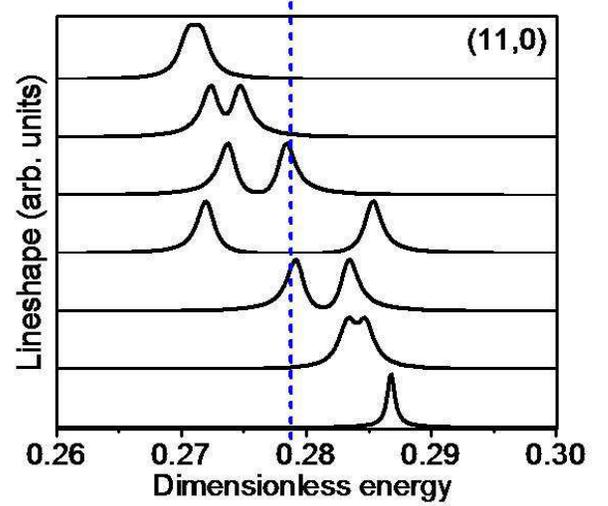}}\caption{(Color
online)~Exciton absorption lineshape as the exciton energy is
tuned to the nearest plasmon resonance [vertical dashed line here;
see Fig.~\ref{fig7}~(a)] in the (11,0) CN. Dimensionless energy is
defined as [\emph{Energy}]/$2\gamma_0$.}\label{fig8}
\end{figure}

Figure~\ref{fig8} shows the calculated exciton absorption
lineshapes as the exciton energy is tuned to the nearest plasmon
resonance of the (11,0) CN. In these calculations, the phonon
relaxation time $\tau_{ph}\!=\!30$~fs was used as reported in
Ref.~\cite{Perebeinos05}.~The line (Rabi) splitting effect is seen
to be $\sim\!0.1$~eV, indicating the strong exciton-plasmon
coupling with the formation of the mixed surface plasmon-exciton
excitations. The splitting is not masked by the exciton-phonon
scattering, and is expected to be larger in smaller diameter
nanotubes.

Obviously, the formation of the strongly coupled mixed
exciton-plasmon states is only possible if the exciton total
energy is in resonance with the energy of an interband surface
plasmon mode.~The exciton energy might be tuned to the nearest
plasmon resonance in ways used for the excitons in semiconductor
quantum microcavities --- thermally (by elevating sample
temperature)~\cite{Reithmaier,Yoshie,Peter}, and/or
electrostatically~\cite{MillerPRL,Miller,Zrenner,Krenner} (via the
quantum confined Stark effect with an external electrostatic field
applied perpendicular to the CN axis).~The two possibilities
influence the different degrees of freedom of the quasi-1D exciton
--- the (longitudinal) kinetic energy and the excitation energy,
respectively [see Eq.~(\ref{Ef})]. In the latter case, in spite of
the fact that the cylindrical surface symmetry of the excitonic
states brings new peculiarities, the general qualitative behavior
of the quantum confined Stark effect in CNs should be similar to
what was previously observed and theoretically analyzed for
semiconductor quantum wells~\cite{MillerPRL,Miller}. One should
expect that the exciton excitation energy $E_{exc}$ and the
interband plasmon energy $E_{p}$ both shift to the red due to the
decrease in the CN band gap $E_g$ as the perpendicular
electrostatic field increases. However, the exciton red shift is
expected to be much less due to the decrease in the absolute value
$|E_b|$ of the exciton (negative) binding
energy~\cite{Bondarev09aps}, which is estimated to be
$\sim\!0.5E_{exc}$ in small-diameter carbon
nanotubes~\cite{Pedersen03,Pedersen04,Wang05,Capaz} and thus
contributes largely to the exciton excitation energy. So, the
exciton excitation energy $E_{exc}$ and the interband plasmon
energy $E_{p}$ will be approaching as the field increases,
bringing the total exciton energy [see Eq.~(\ref{Ef})] in
resonance with the plasmon mode due to the non-zero longitudinal
kinetic energy term at finite temperature.

\begin{figure}[t]
\epsfxsize=8.65cm\centering{\epsfbox{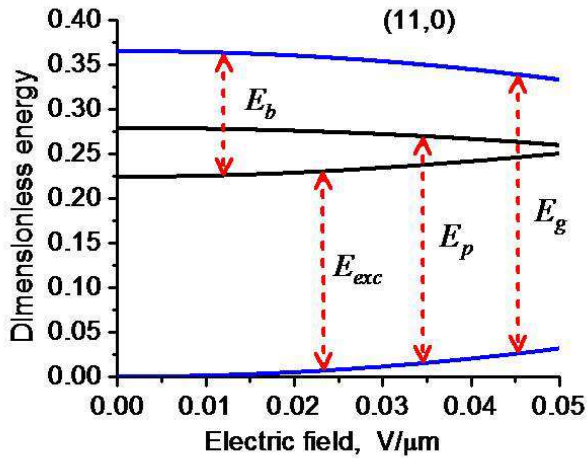}}\caption{(Color
online)~Calculated dependences of the first bright exciton
parameters in the (11,0) CN on the electrostatic field applied
perpendicular to the CN axis. Dimensionless energy is defined as
[\emph{Energy}]/$2\gamma_0$. See text for notations.}\label{fig9}
\end{figure}

Figure~\ref{fig9} shows the results of the recent calculations of
the quantum confined Stark effect for the (11,0)
nanotube~\cite{Bondarev09aps}, which confirm the qualitative
expectations just discussed.~More details on these calculations
and the complete theory of the quantum confined Stark effect in
CNs will be published elsewhere.

~

\textbf{5. Conclusion}

~

Similar to semiconductor microcavities and photonic band-gap
materials, carbon nanotubes may qualitatively change the character
of the atom-electromagnetic-field interaction, yielding the
\emph{strong} atom-field coupling. For atoms (ions) doped into
small-diameter ($\sim\!1$~nm) nanotubes, the effect of the optical
absorption line splitting (Rabi splitting) of $\sim\!1$~eV is
predicted in the vicinity of the energy of the atomic transition.
This is at least three orders of magnitude larger than the Rabi
splitting of $\sim\!140\!-\!400$~$\mu$eV observed recently in the
photoluminescence experiments for exciton-polaritons in quantum
dots in semiconductor
microcavities~\cite{Reithmaier,Yoshie,Peter}. The larger effect in
atomically doped CNs comes from the facts that (i)~typical atomic
binding energies are at least three orders of magnitude larger
than typical excitonic binding energies in solids, and (ii)~the
effective mode volume of the medium-assisted vacuum-type
electromagnetic field in the CNs is small enough to achieve Rabi
splitting of $\sim\!1$~eV in 1~nm-diameter nanotubes.

Rear-earth Eu$^{3+}$ complexes in hydrocarbon solvents might be
good candidates to test this prediction. They are known to be
excellent probes to study quantum optics phenomena in spatially
confined systems such as dielectric microspheres~\cite{Schniepp},
or photonic crystals~\cite{Gaponenko}, owing to the dominant
narrow, easily detectible $^{5}D_{0}\!\rightarrow\!^{7}F_{2}$
dipole transition between two deep-lying electronic levels of
europium that essentially create a two-level system.

Metallic carbon nanotubes of $\sim\!1$~nm diameter are
demonstrated theoretically to be very efficient, even at room
temperatures, to entangle a pair of the spatially separated atomic
qubits located in the middle of the nanotube, provided they
interact with the same resonance of nanotube's local photonic DOS.
This latter condition is automatically satisfied for identical
atoms doped into the nanotube. The entanglement greatly exceeds
0.35 (known to be the maximal value one can achieve in the weak
atom-field coupling regime~\cite{Dung}), and persists with no
damping for very long times. Such a behavior is attributed to the
strong atom-field coupling and electronic structure resulting in
the resonator-like distance dependence of the two-particle local
photonic DOS in metallic CNs. Thus, similar to quasi-0D excitonic
polaritons in quantum dots in semiconductor microcavities
suggested recently to be a possible way to produce the excitonic
qubit entanglement~\cite{Hughes}, small-diameter metallic
atomically doped carbon nanotubes might represent a novel approach
towards solid-state quantum information transfer over long
distances --- centimeter-long distances, as a matter of fact,
since centimeter-long small-diameter single-walled CNs are
currently technologically available~\cite{Zheng,Huang}.

In the current version of the theory outlined here, all the
possible decoherence mechanisms in the two-qubit system coupled to
nanotube's surface electromagnetic mode are taken into account in
terms of the relaxation time approximation in the CN surface
conductivity. The conductivity appears in the photonic DOS and
plays a crucial role in the entanglement of formation. It is also
important that the method of entanglement preparation described
here does not involve the spin degrees of freedom --- just a
normal, but strong, electric dipole coupling to the same local
(single-particle) photonic DOS resonance of the nanotube is
required to entangle the pair of spatially separated atomic
qubits, as opposed to the spin entanglement studied recently in
other fullerene-based models~\cite{Benjamin}. The outlined
entanglement scheme can be further generalized to the multi-atom
entanglement via the nearest neighbor pairwise quantum
correlations, thus challenging novel applications of atomically
doped CNs in quantum information science.

In pristine small-diameter ($\lesssim\!1$~nm) semiconducting CNs,
the strong exciton-surface-plasmon coupling effect is demonstrated
with the characteristic exciton absorption line (Rabi) splitting
$\sim\!0.1$~eV.~This is almost as large as the typical exciton
binding energies in such CNs
($\sim\!0.3\!-\!0.8$~eV~\cite{Pedersen03,Pedersen04,Wang05,Capaz}),
and of the same order of magnitude as the exciton-plasmon Rabi
splitting in organic semiconductors
($\sim\!180$~meV~\cite{Bellessa}).~Also, this is much larger than
the exciton-polariton Rabi splitting in semiconductor
microcavities
($\sim\!140-400\,\mu\mbox{eV}\,$\cite{Reithmaier,Yoshie,Peter}),
or the exciton-plasmon Rabi splitting in hybrid
semiconductor-metal nanoparticle molecules~\cite{Govorov}. Also
important is that the strong exciton-plasmon coupling effect takes
place in an individual carbon nanotube as opposed to artificially
fabricated hybrid plasmonic nanostructures just mentioned. To
bring the exciton in resonance with nanotube's nearest plasmon
mode, the exciton energy can be tuned in ways used for the
excitons in semiconductor quantum microcavities --- thermally (by
elevating sample temperature)~\cite{Reithmaier,Yoshie,Peter},
and/or electrostatically~\cite{MillerPRL,Miller,Zrenner,Krenner}
(via the quantum confined Stark effect with an external
electrostatic field applied perpendicular to the CN axis). The two
possibilities affect the different degrees of freedom of the
quasi-1D excitons in CNs --- the (longitudinal) kinetic energy and
the excitation energy, respectively.

To conclude, the surface electromagnetic field phenomena reviewed
here have a great potential to open up new paths for the
development of the CN based tunable optoelectronic device
applications in areas such as nanophotonics, nanoplasmonics,
cavity quantum electrodynamics, and quantum information science.

\section{Acknowledgements}

\noindent This work was supported by the US National Science
Foundation (grants ECS-0631347 and HRD-0833184), the US Department
of Defense (grant W911NF-05-1-0502) and NASA (grant NAG3-804).
Discussions with M.A.Braun, J.Finley, S.V.Gaponenko, M.F.Gelin,
A.O.Govorov, A.Kleinhammes, J.Liu, G.E.Malashkevich, H.Qasmi,
B.Vlahovic, L.Woods, and Y.Wu are gratefully acknowledged.

\end{document}